\documentclass[aps,nofootinbib,reprint,nobibnotes,showpacs,superscriptaddress]{revtex4-1} 

\usepackage[english]{babel}
   \addto\captionsenglish{
                         }

\usepackage{graphicx}
   \graphicspath{{graphics/}}

\usepackage[caption=false]{subfig}

\usepackage[shortlabels]{enumitem}

\usepackage{amsmath}
\usepackage{amssymb}
\usepackage[version=3]{mhchem}
\usepackage{mathrsfs}
\usepackage{bm}
\usepackage{multirow}

\usepackage[hidelinks,linktocpage]{hyperref}
\hypersetup{
            colorlinks,
            linkcolor={red!70!black},
            citecolor={green!70!black},
            urlcolor={blue!70!black}
           }

\usepackage[usenames,dvipsnames]{xcolor}
\usepackage{comment} 
\usepackage{slashed} 
\usepackage{cancel}
\usepackage[normalem]{ulem}
\usepackage{lineno}

\newcommand{\bb}{\ensuremath{0\nu\beta\beta}} 
\newcommand{\TeO}{TeO\ensuremath{_2}}
\newcommand{\Tecxx}{\ensuremath{^{120}}Te}
\newcommand{\Tecxxx}{\ensuremath{^{130}}Te}
\newcommand{\Co}{\ensuremath{^{60}}Co}
\newcommand{\Bi}{\ensuremath{^{214}}Bi}
\newcommand{\bEC}{\ensuremath{\beta^+ EC}}
\newcommand{\nnbEC}{\ensuremath{0 \nu \beta^+ EC}}
\newcommand{\ckky}{\mbox{\ensuremath{\textrm{counts}/(\textrm{keV}\cdot\textrm{kg}\cdot\textrm{yr}})}}
\newcommand{\cky}{\mbox{\ensuremath{\textrm{counts}/(\textrm{kg}\cdot\textrm{yr}})}}

\newcommand\jo[2][]{{\color{blue} \ifx#1&\else\sout{#1} \fi #2}}

\newcommand\ali[2][]{{\color{red} \ifx#1&\else\sout{#1} \fi #2}}

\begin{document}

 \title{Search for Neutrinoless \bEC\ Decay of \Tecxx\ with CUORE}

\author{D.~Q.~Adams}
\affiliation{Department of Physics and Astronomy, University of South Carolina, Columbia, SC 29208, USA}

\author{C.~Alduino}
\affiliation{Department of Physics and Astronomy, University of South Carolina, Columbia, SC 29208, USA}

\author{K.~Alfonso}
\affiliation{Department of Physics and Astronomy, University of California, Los Angeles, CA 90095, USA}

\author{F.~T.~Avignone~III}
\affiliation{Department of Physics and Astronomy, University of South Carolina, Columbia, SC 29208, USA}

\author{O.~Azzolini}
\affiliation{INFN -- Laboratori Nazionali di Legnaro, Legnaro (Padova) I-35020, Italy}

\author{G.~Bari}
\affiliation{INFN -- Sezione di Bologna, Bologna I-40127, Italy}

\author{F.~Bellini}
\affiliation{Dipartimento di Fisica, Sapienza Universit\`{a} di Roma, Roma I-00185, Italy}
\affiliation{INFN -- Sezione di Roma, Roma I-00185, Italy}

\author{G.~Benato}
\affiliation{INFN -- Laboratori Nazionali del Gran Sasso, Assergi (L'Aquila) I-67100, Italy}

\author{M.~Beretta}
\affiliation{Department of Physics, University of California, Berkeley, CA 94720, USA}

\author{M.~Biassoni}
\affiliation{INFN -- Sezione di Milano Bicocca, Milano I-20126, Italy}

% ...
\date{\today}

 \begin{abstract}
CUORE is a large scale cryogenic experiment searching for neutrinoless double beta decay (\bb) in \Tecxxx. The CUORE detector is made of natural tellurium, providing the possibility of rare event searches on isotopes other than \Tecxxx. In this work we describe a search for neutrinoless positron emitting electron capture (\bEC) decay in \Tecxx\ with a total \TeO\ exposure of 355.7 kg $\cdot$ yr, corresponding to 0.2405 kg $\cdot$ yr of \Tecxx. 
Albeit $0 \nu \beta \beta$ with two final state electrons represents the most promising channel, the emission of a positron and two 511-keV $\gamma$s make \nnbEC\ decay signature extremely clear. To fully exploit the potential offered by the detector modularity we include events with different topology 
and perform a simultaneous fit of five selected signal signatures. Using blinded data we extract a median exclusion sensitivity of $3.4 \cdot 10^{22}$ yr at 90\% Credibility Interval (C.I.). After unblinding we find no evidence of \nnbEC\ signal and set a 90\% C.I. Bayesian lower limit of $2.9 \cdot 10^{22}$ yr on \Tecxx\ half-life. 
This result improves by an order of magnitude the existing limit from the combined analysis of CUORE-0 and Cuoricino. \\[10pt]
Published on: Physical Review C 105,065504 (2022) \hfill DOI:~\href{https://link.aps.org/doi/10.1103/PhysRevC.105.065504}{10.1103/PhysRevC.105.065504}
 \end{abstract}
 \maketitle

 \section{Introduction}
 \label{sec:Intro}
The quest to understand the nature of the neutrino mass has launched a world-wide effort to search for neutrinoless double beta (\bb) decay~\cite{Agostini:2021kba,Dolinski:2019nrj}. The observation of this lepton number violating ($\Delta L = 2$) decay would conclusively demonstrate that neutrinos are Majorana fermions (i.e. their own antiparticles) and provide further evidence for the role of physics beyond the Standard Model (SM).
 
Double beta decay 
is a spontaneous weak process changing the nuclear charge $Z$ by two units while leaving the atomic mass $A$ unchanged. At present, the most studied mechanism is the $\beta^-\beta^-$ decay that features the emission of two electrons. However, depending on the relative number of protons and neutrons in a nucleus, three additional processes are possible~\cite{Blaum:2020ogl,Barea:2013wga,Kotila:2013gea}: double electron capture ($EC EC$), double positron decay ($\beta^+ \beta^+$) and positron emitting electron capture (\bEC). $ECEC$ is preferred by the available phase-space, but the rate is typically reduced by several orders of magnitude because an extra radiative process is required to satisfy energy-momentum conservation~\cite{Doi:1993nmd}.
$\beta^+ \beta^+$ and \bEC\ have a clear signature due to the presence of positrons in the final state. 
Furthermore, the \bEC\ mode shows an enhanced sensitivity to right-handed weak currents~\cite{Hirsch:1994nsc} and could play an important role in the comprehension of the underlying mechanism in the event of a \bb\ discovery. The first direct observation 
of two-neutrino $ECEC$ decay was made in $^{124}$Xe with the XENON1T detector~\cite{Aprile:2019dti}. Half-life estimates for \nnbEC\ in the most promising nuclei are of the order of 10$^{29}$-10$^{33}$ yr (for $\langle m_{\nu} \rangle$ = 20\,meV)~\cite{Barea:2013wga}, experimental limits are in the $10^{18}$-$10^{21}$ range in the isotopes studied:  $^{74}$Se~\cite{Barabash:2020fnn}, $^{64}$Zn~\cite{Belli:2008zzd}, $^{112}$Sn~\cite{Barabash:2008wj} and $^{120}$Te~\cite{Alduino:2017dbf,Dawson:2009cc,Barabash:2007iu}. 

In this work, we describe a search for neutrinoless positron emitting electron capture decay (\nnbEC) of \Tecxx\ with CUORE, an array of \TeO\ crystals operated as cryogenic detectors at the Laboratori Nazionali del Gran Sasso in central Italy. The primary goal of the experiment is the search for \bb\ decay of \Tecxxx~\cite{Artusa:2014lgv}, but the use of tellurium with natural isotopic composition allows us to search for other rare decays~\cite{Campani:2021cfh}, such as double beta decay in $^{128}$Te~\cite{Adams:2022} and, indeed, \Tecxx. 

\Tecxx\ has a natural isotopic abundance of 0.09(1)\%~\cite{Meija:2016IUPAC} and could potentially decay to $^{120}$Sn via $0 \nu EC EC$ and \nnbEC. We did not investigate the former channel as this mode is expected to be suppressed except in the case of the existence of a resonance condition~\cite{Blaum:2020ogl}. At present the most stringent limit on \Tecxx\ \nnbEC\ half life is $T_{1/2} > 2.7 \cdot 10^{21}$ yr (90\% C.I.) and was obtained with the combination of the CUORE-0 and Cuoricino results~\cite{Alduino:2017dbf}. A limit of $T_{1/2}> 7.6 \cdot 10^{19}$ yr at 90\% C.L. on the $2\nu$ mode was set with Cuoricino data~\cite{Andreotti:2010nn}.

To date \Tecxx\ has been minimally investigated from a theoretical point of view and no calculation of the nuclear matrix elements is available for its decay. Improved values of the phase space factors for both the $2 \nu$ mode and the $0 \nu$ mechanism are reported in \cite{Kotila:2013gea}, while \cite{Abad:jpa-00224042} gives an estimate for the half life of $2\nu$\bEC\ decay of $T_{1/2} = 4.4 \cdot 10^{26}$ yr.

The \nnbEC\ decay of \Tecxx\ can be expressed as:
\begin{equation}
\label{eq:decay-reaction}
\begin{split}
{}^{120} \mathrm{Te} + e^{-}_{b} &\rightarrow {}^{120} \mathrm{Sn}^* + \beta^{+} \\
&\rightarrow {}^{120} \mathrm{Sn} + X +  \beta^{+} \\
&\rightarrow {}^{120} \mathrm{Sn} + X +  2 \gamma_{511} \\
\end{split}
\end{equation}
where $e^{-}_{b}$ indicates the electron captured from an atomic shell with binding energy $\mathrm{E}_b$ and $X$ indicates an Auger electron or an X-ray emitted in the $^{120}$Sn de-excitation. The back-to-back 511 keV $\gamma$ rays are the product of $e^+-e^-$ annihilation. Given the absence of neutrinos, the available energy is shared between the four particles in the final state, with the daughter nucleus being almost at rest because of its larger mass. In literature, we find only one direct measurement of the Q-value, i.e. the difference between \Tecxx\ and $^{120}$Sn atomic masses, obtained with a Penning trap~\cite{Scielzo:2009nh}, $Q = (1714.8 \pm 1.3)$~keV. The emitted positron has a kinetic energy of 
\begin{equation}
\label{eq:kinetic-positron}
K = Q- 2 m_e c^2 - E_b \ , 
\end{equation}
where $E_b$ indicates the binding energy of the original atomic shell. In the likely assumption of electron capture from the K-shell, $E_b$ is 29.2~keV\footnote{The binding energies for the L-shell are 4.46, 4.15 and 3.93~keV, i.e. 2s, 2p$_{1/2}$, 2p$_{3/2}$ levels. The ratio of L-capture to K-capture is 12\% for the $^{120}$Sb $\rightarrow$ $^{120}$Sn $EC$ decay.}~\cite{X-rayDataBooklet} and $K$ is 663.6~keV.

The analysis presented here exploits the granularity of the CUORE detector to reconstruct event topologies via a coincidence analysis, thereby minimizing contributions from background sources and optimizing our sensitivity to \nnbEC\ decay. We select the topological signatures with the best signal-to-background ratio, and use five of the six signatures used in the joint Cuoricino/CUORE-0 analysis~\cite{Alduino:2017dbf}. 
\section{The CUORE experiment}
\label{sec:CUORE-Analysis}
The Cryogenic Underground Observatory for Rare Events (CUORE) is an underground tonne-scale experiment designed to search for neutrinoless double beta decay in \Tecxxx. The detector is a close-packed array of 988 \TeO\ crystals operated as calorimeters at a cryogenic temperature of 10 mK. The temperature is maintained by means of a custom-made cryogen free $^3$He-$^4$He dilution refrigerator~\cite{Alduino:2019xia}. The crystals are arranged in a cylindrical matrix of 19 identical towers. Each tower hosts 52 $5 \times 5 \times 5$ cm$^3$ cubic detectors divided in 13 floors of 4 modules each. Every crystal weighs $\sim$750\,g 
for a total \TeO\ mass of 742\,kg, corresponding to 
0.5\,kg of \Tecxx. Several shields are employed to protect the calorimeters from external $\gamma$s, neutrons, and the radioactive background coming from the cryogenic infrastructure itself~\cite{Alduino:2019xia,Pattavina:2019pxw,Alduino:2017ztm}. 
Any energy deposition, e.g. following a $\beta$ or $\alpha$ decay, causes an increase in the crystal's temperature that is measured with a neutron transmuted doped (NTD) germanium thermistor~\cite{Larrabee:1984} glued directly to the crystal surface.

The data acquisition and production chain follows closely the same strategy outlined in \cite{Adams:2019jhp}. Here we discuss the main steps required to convert the raw thermal pulses into an energy spectrum of \nnbEC\ candidate events, highlighting differences with respect to the \Tecxxx\ \bb\ decay analysis. The voltage across the NTD thermistor of each crystal is amplified, filtered through a 6-pole Bessel anti-aliasing filter, and digitized with a sampling frequency of 1\,kHz~\cite{DiDomizio:2018ldc,Arnaboldi:2017aek}. During data acquisition, we save continuous detector waveforms that are digitally triggered offline. To increase the signal-to-noise ratio (SNR) and study low energy phenomena, we apply a low threshold trigger algorithm based on the optimum filter (OF) technique~\cite{Campani:2020ltd,Alduino:2017xpk}. We divide our data into time periods of one to two months characterized by the same operating conditions and refer to them collectively as \emph{datasets}. A calibration period of few days marks the end of a dataset and the start of the following one. The data collected in between are called \emph{physics data} and used for double beta decay searches. 

For each triggered pulse, we analyze a 10-s window consisting of 3\,s before and 7\,s after the trigger time. The pre-trigger voltage serves as a proxy for the temperature at the time of the event, while the pulse amplitude indicates the energy absorbed by the crystal. To improve the SNR, we build the OF transfer function of each calorimeter from a signal pulse template and the measured noise power spectrum. We filter the raw event waveform and apply a correction against changes in gain caused by slow drifts in the detectors' temperatures. We reconstruct the energy using the most intense gamma lines of $^{232}$Th and $^{60}$Co as calibration sources~\cite{Adams:2019jhp}. The calibration function is a second-order polynomial with the intercept constrained to be zero. We remove periods of time with sub-optimal detector performance or where the processing failed, and apply a set of basic quality cuts to reject events with poor energy reconstruction or affected by pile-up, i.e. secondary pulses within the same window of the main event. 
Finally, we combine the information extracted from six pulse shape parameters to reject noisy events, pile-up, or other spurious events that survived the previous cuts based on their degree of resemblance to a clean sample of particle events~\cite{Adams:2019jhp}. 

\subsection{Data selection}
\label{subsec:CUORE-data}

The analysis presented here is based on the same set of data (divided into 7 datasets) described in \cite{Adams:2019jhp} with two major differences. 
First, we exclude a subset of the calorimeter-dataset pairs that have sub-optimal performance over the $400 - 1800$\,keV range. These were included in the \bb\ decay analysis which focuses only on the region of interest (ROI) around 2528\,keV, but were not suitable for the present one. The final \TeO\ exposure 
amounts to 355.7 kg $\cdot $ yr, corresponding to 0.2405 kg $\cdot$ yr of \Tecxx. Second, while the \bb\ decay analysis uses an \emph{anti-}coincidence cut to veto non-signal-like events, the present analysis \emph{requires} specific coincidence conditions in order to select signal-like events. These are defined explicitly in Sec.~\ref{subsec:signatures}. We define a coincidence as a simultaneous energy deposition in two or more crystals. Specifically, we require at least 70\,keV of energy to be released in each crystal within a 30\,ms time window, and only apply the reconstruction to crystals that are at most 15\,cm apart from each other. Simultaneous events in two calorimeters are said to be \emph{multiplicity} 2 ($\mathcal{M}_2$). Coincidences can be chained together to form higher multiplicities of $n$ coincident events ($\mathcal{M}_n$). Most events are not in coincidence and are simply called multiplicity 1 ($\mathcal{M}_1$).

\section{\nnbEC\ Decay Search}
\label{sec:Analysis}
\subsection{Decay signatures}
\label{subsec:signatures}
\begin{table*}[hbt]
\centering
  \caption{Selected experimental signatures of \Tecxx\ \nnbEC\ decay in CUORE. For each signature we list the final state particles detected, the expected energy of the \nnbEC\ signal, the multiplicity, 
  and the fit range(s) $\Delta \mathrm{E}_i$, with $i = 0,$ ..., $M-1$. As outlined in Sec.~\ref{subsec:MC-efficiency}, the last column shows the containment efficiency $\varepsilon_{\mathrm{mc}}$ with the uncertainty reported in round brackets.}
  \label{tab:120TeS_properties}
     \begin{tabular}{cccccccc}
      \toprule
      & Particles & Signal Peak &  & \multicolumn{3}{c}{Energy range [keV]} & Containment efficiency \\\cline{5-7}
       Signature & Detected & Position [keV] & Multiplicity & $\Delta \mathrm{E}_0$ & $\Delta \mathrm{E}_1$ & $\Delta \mathrm{E}_2$ & $\varepsilon_{\mathrm{mc}}$ [\%] \\
      \colrule
(a) & $\beta^++X+\gamma_{511}$ & 1203.8 & 1 & [1150,1250] & & & 12.8(5) \\ 
(b) & $\beta^++X+2\gamma_{511}$ & 1714.8 & 1 & [1703,1775] & & & 13.1(5) \\ 
(c) & $\big(\beta^++X,\ \gamma_{511}\big)$ & (692.8, 511) & 2 & [650,750] & [460,560] & & 4.10(20) \\ 
(d) & $\big(\beta^++X+\gamma_{511},\ \gamma_{511}\big)$ & (1203.8, 511) & 2 & [1150,1250] & [460,560] & & 13.8(6) \\ 
(e) & $\big(\beta^++X, \ \gamma_{511}, \ \gamma_{511} \big)$ & (692.8, 511, 511) & 3 & [650,750] & [460,560] & [460,560] & 2.15(9) \\ 
           \toprule
    \end{tabular}
\end{table*}
This section describes the experimental signatures of \nnbEC\ decay in CUORE. Eq.~\ref{eq:decay-reaction} shows that the available energy in such a transition is shared between the four final state particles. In this analysis, we focus on events that satisfy the following conditions:
\begin{enumerate}
    \item the $^{120}$Sn daughter nucleus deposits its recoil energy in the source \nnbEC\ crystal, i.e. the crystal where the decay occurred;
    \item the product $X$ of $^{120}$Sn de-excitation, i.e. an X-ray or Auger electron, is fully absorbed by the source \nnbEC\ crystal; 
    \item \label{item:positron_containment} the positron deposits its kinetic energy and annihilates in the source \nnbEC\ crystal; 
   \item \label{item:gamma_containment} each 511\,keV $\gamma$ is either fully absorbed by a single crystal (either originating or neighboring) or fully escapes the active part of the detector. 
\end{enumerate}
In practice, we are selecting events where each final state particle fully releases its energy in no more than one crystal. We therefore ignore events in which a Compton scattering splits the energy of a single gamma among two or more detectors. This requirement greatly simplifies the experimental signatures at a cost to the signal efficiency, especially due to requirement \ref{item:gamma_containment}. 

These conditions limit to three the maximum number of crystals simultaneously involved in a \nnbEC\ event. 
Six experimental signatures are possible: three with $\mathcal{M}_1$ events, two with $\mathcal{M}_2$ events and one including $\mathcal{M}_3$ events; however, as discussed below, we exclude one of the $\mathcal{M}_1$ signatures. The $\mathcal{M}_1$ signatures are those in which all the energy is deposited in a single crystal or in passive materials. If both gammas escape the crystal, we expect a peak at $K+E_b=692.8$\,keV; if only one gamma escapes, we have $Q-m_e c^2=1203$\,keV; if none of them escapes, we expect a peak at $Q=1714.8$\,keV. For the $\mathcal{M}_2$ signatures, we expect 511\,keV in one crystal and 692.8 or 1203.8\,keV in the other one, depending on whether the second 511\,keV gamma escapes or not.
The $\mathcal{M}_3$ signature includes events in which one crystal absorbs the positron and the $^{120}$Sn de-excitation products, while two neighboring crystals see one 511\,keV $\gamma$ each. In order to minimize the contribution from background we impose a topological requirement on \nnbEC\ candidates in $\mathcal{M}_3$: the crystals with the two 511\,keV $\gamma$s must be on opposite sides of the originating crystal, 
because the annihilation $\gamma$s are produced back-to-back.

\begin{figure*}[htpb]
    \centering
    \hspace{-0.5cm} 
    \includegraphics[width=0.4\textwidth]{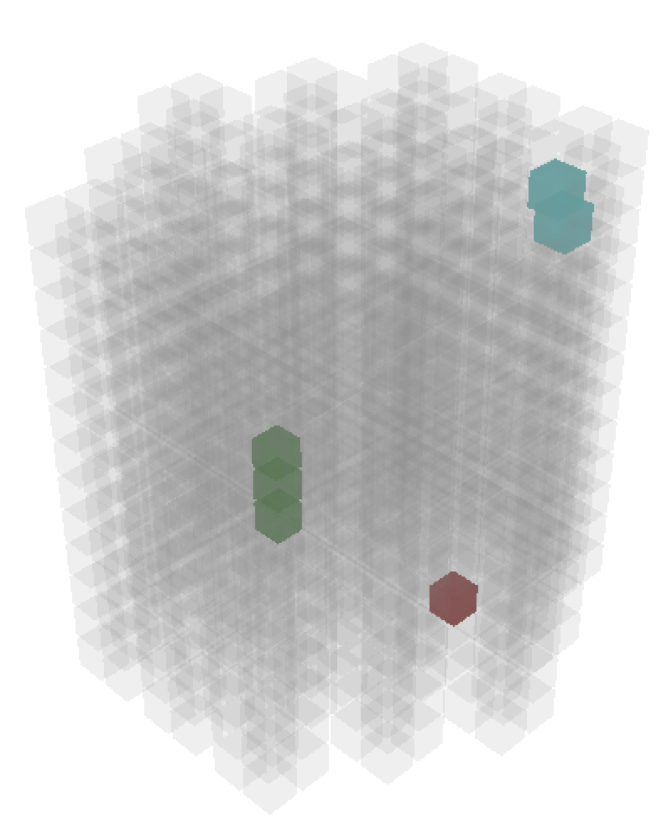} \hspace{+0.1cm} 
    \includegraphics[width=0.46\textwidth]{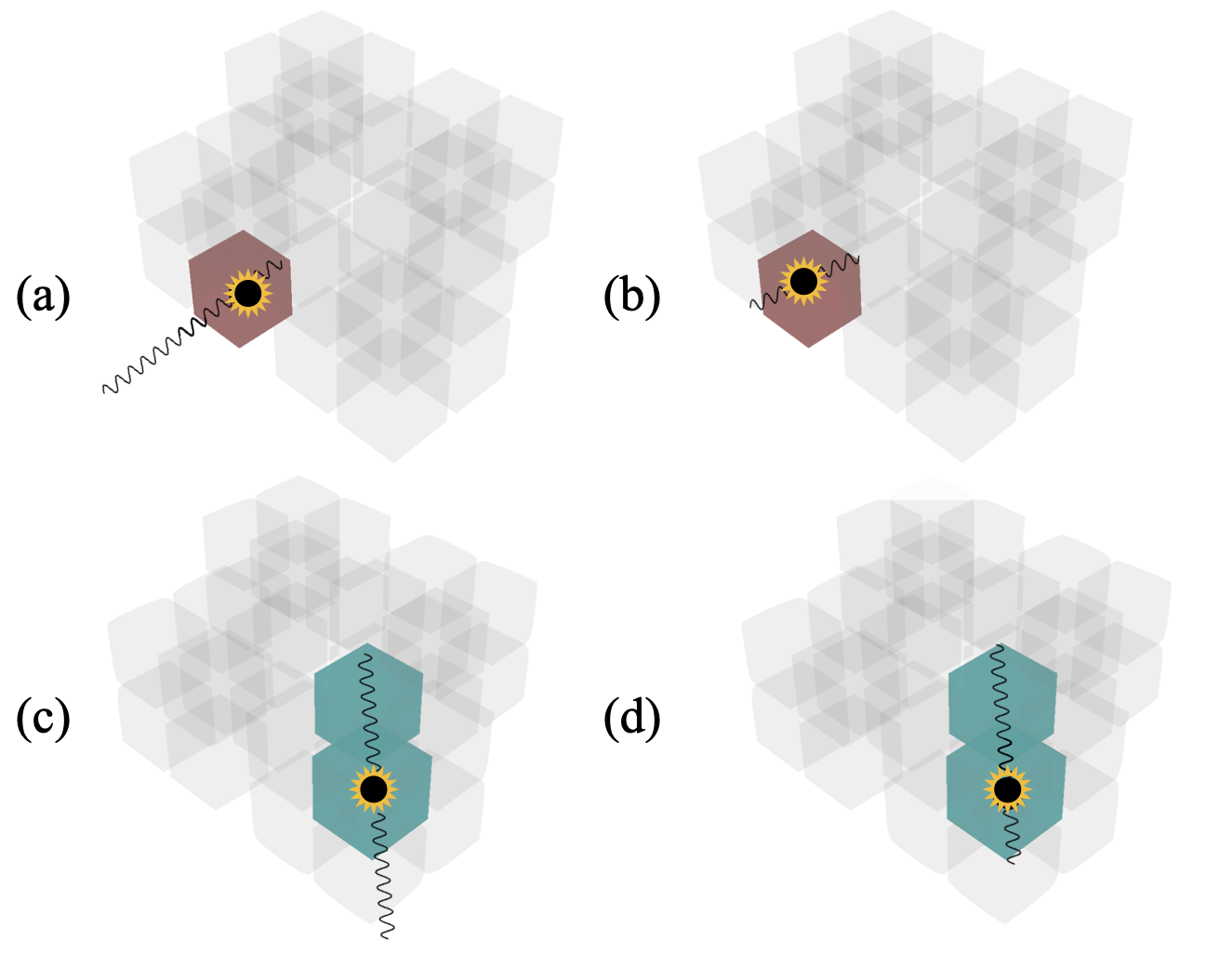} \hspace{+0.1cm}
    \includegraphics[width=0.1\textwidth]{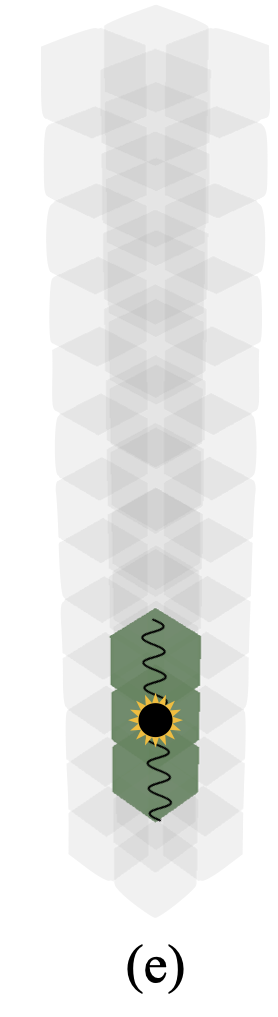} 
    \caption{(Left) Signatures of \Tecxx\ \nnbEC\ decay in CUORE: examples of $\mathcal{M}_1$, $\mathcal{M}_2$ and $\mathcal{M}_3$ topologies are highlighted; each translucent gray cube represents an individual CUORE crystal. (Right) Specific topology of each signature included in this analysis. The crystal where the decay occurs has a symbol overlaid, whereas curly lines represent the 511-keV $\gamma$s following $\beta^+$ annihilation. Signature (a) and (c) feature a gamma escape, all the others are full-containment signatures. Signatures (c)-(e) include a distance cut of 15 cm on the involved crystals. 
    $\mathcal{M}_3$ events must reflect the fact that the 511\,keV gammas are emitted back-to-back. 
    }
    \label{fig:signatures}
\end{figure*}

We decided to exclude the $\mathcal{M}_1$ signature centered at $692.8$\,keV since we expect this signature to add only a negligible contribution to the overall sensitivity of our study: the sensitivity scales as $\varepsilon/\sqrt{b}$ in the case of non-negligible background and here the signal efficiency is 1.4\%, i.e. the smallest if compared with Table~\ref{tab:120TeS_properties}, and the background index is $\sim$2\,\ckky, i.e. the highest considering Tab.~\ref{tab:fitModel120Te}.
The selected signatures of \Tecxx\ \nnbEC\ decay in CUORE are illustrated in Fig.~\ref{fig:signatures}. Their properties are summarized in Table~\ref{tab:120TeS_properties}. 
We label them with alphabetical letters: (a) and (b) include $\mathcal{M}_1$ events, (c) and (d) are $\mathcal{M}_2$ signatures and (e) features $\mathcal{M}_3$ events. For each signature we select the widest possible fit ranges 
that allow us to constrain the background rate 
without introducing unnecessary peaks or structures into the analysis. In this respect, the range of signature (b) is narrower than the other ones to avoid including a potential $^{214}$Bi peak at 1693 keV. 

\subsection{Decay simulations and containment efficiency} 
\label{subsec:MC-efficiency}
We evaluate the containment efficiencies, i.e. the probability that 
any given \nnbEC\ decay matches one of the signature requirements, by means of a Monte Carlo simulation using the standard CUORE MC framework~\cite{Adams:2020bok,Alduino:2017ztm} based on GEANT4~\cite{Agostinelli:2003nim}.

We simulate \nnbEC\ by generating the particles emitted in the decay as primaries. In the \nnbEC\ process the nucleus captures one atomic shell electron, most likely from the K-shell, and simultaneously converts two protons to neutrons and emits a positron. The initial electron capture can occur from a variety of electron shells and then emit either an X-ray or Auger electron. For an atomic shell with binding energy $E_b$, the kinetic energy of the emitted positron is given by Eq.~\ref{eq:kinetic-positron}. The way different electron shells contribute to the transition is fixed by the atomic properties of the material as stated in Sec.~\ref{sec:Intro}. This leads to a complex list of scenarios that must be simulated and combined with correct weights. Instead, we take a simplified and overly-conservative approach and consider the worst case scenario in which all the final state particles have the maximum chance of escape, i.e. the positron is ejected with maximal kinetic energy ($K = Q-2m_e c^2=692.8$\,keV, corresponding to a negligible binding energy) and in the atomic relaxation a K-shell X-ray of energy $E_b=29.2$\,keV is emitted, even if these situations are clearly mutually exclusive. With this approach we can extract a lower limit to the real containment efficiency.  
We find that the effect on the final result is small enough to justify the simplification. We determine the containment efficiency for a 29.2\,keV X-ray to be 99.7\% by simulating $10^{6}$ primaries uniformly distributed over the crystals, and include this as part of the signal efficiency. Next, we generate 10$^7$ 692.8\,keV positrons as primary particles uniformly distributed in the \TeO\ volume and evaluate the \nnbEC\ containment efficiency using the same coincidence selection cuts described in Sec.~\ref{subsec:CUORE-data}.       

The procedure to evaluate the containment efficiency slightly differs for $\mathcal{M}_1$ vs. higher-multiplicity signatures. In all cases, we discard events in which one or more of the final state particles release only a fraction of their energy in a single crystal. In other words, we select only events at the nominal energies for each signature.
In the $\mathcal{M}_1$ case, we select all the events that lie within the energy ranges listed in Table~\ref{tab:120TeS_properties} and we define a suitable fitting model, namely a Gaussian peak for the \Tecxx\ \nnbEC\ signal plus additional terms to parameterize the continuum~\cite{Campani_PhD-thesis:2021}. We then compute the containment efficiency as the ratio of the number of events populating the \nnbEC\ peak (i.e. the integral of the fitted Gaussian) over the number of generated decays.
For $\mathcal{M}>1$ signatures, we select all the particle events which satisfy one of the cuts specified in Table~\ref{tab:120TeS_properties}, project them onto the directions of the 511-keV $\gamma$s (either one or two depending on the specific signature) and perform a Gaussian fit to further constrain the selection around the $\gamma$ peaks. 
This guarantees that the MC efficiency is evaluated only on events belonging to the $\mathcal{M}$-dimensional signal peak. Finally, we project the surviving events onto the $\beta^+$ axis and extract the containment efficiency from a peak+continuum fit as for the $\mathcal{M}_1$ signatures. The results are reported in the last column of Table~\ref{tab:120TeS_properties}: the total containment efficiency is $\sim$46\%, with other contributions coming either from partial energy depositions in passive components or higher-multiplicity events ($\sim$5\%). The containment efficiency has two sources of uncertainty: the fit model 
used to compute the number of \nnbEC\ decay events, that has negligible effects ($\lesssim0.1\%$), and the models used in GEANT4 to reproduce gamma rays interactions, in particular Compton scattering~\cite{Allison:2016lfl}. We take the relative difference between the Compton scattering attenuation coefficient evaluated with several GEANT4 models and reference data~\cite{Allison:2016lfl} as a measure of the relative uncertainty on this efficiency term and set a 4\% effect on the containment efficiency of all signatures. 
\subsection{Detection efficiency} 
\label{subsec:efficiencies}
The detection efficiency of a given \nnbEC\ signature is the product of three terms: the containment efficiency ($\varepsilon_{mc}$), the
analysis cut efficiency ($\varepsilon_{\rm cut}$), and the probability of tagging events with the correct multiplicity ($\varepsilon_{\rm ac}$). Both $\varepsilon_{\rm cut}$ and $\varepsilon_{\rm ac}$ are evaluated at dataset level.

The analysis cut efficiency combines the effects of basic quality cuts and pulse shape analysis (PSA) on the final event selection. We closely follow the procedure outlined in \cite{Adams:2020bok} to evaluate both efficiency terms. The base cut efficiency is the product of detection (trigger), energy reconstruction and pile-up rejection efficiency. We use injected heater pulses~\cite{Andreotti:2012zz} to evaluate the fraction of events correctly flagged by the trigger algorithm and whose energy is properly reconstructed, as well as the probability of false positives in the identification of pile-up events. Given the large statistics available, we estimate all the contributions separately for each crystal and average them over the dataset. On the contrary, PSA efficiency is extracted for an entire dataset from the survival probability of two independent samples: $\mathcal{M}_2$ events whose total energy is compatible with $\gamma$ lines from known background sources, and single crystal events corresponding to fully absorbed $\gamma$ lines, such as those from $^{40}$K and $^{60}$Co. The first sample includes events in a wide range of energies, and allows the evaluation of the PSA efficiency as a function of energy. The second has higher statistics but only at a small set of fixed energies, rather than on a continuum. For the sake of this analysis, the PSA efficiency does not depend on energy and can be treated as a constant. We define it as the average of the values obtained from the two samples and include it in the analysis cut efficiency $\varepsilon_{\rm cut}$. We treat the difference between the two methods 
as a systematic effect, adding a scaling parameter common to all datasets in the final fit ($\pm0.7$\%). 

Every time a particle releases its energy in a crystal, there is a small but non-zero probability that a completely uncorrelated event occurs nearly simultaneously in some other channel. The probability that an event is correctly categorized in terms of multiplicity is called the anti-coincidence efficiency, and evaluated using the $^{40}$K 1461\,keV $\gamma$ emission, which is expected to reconstruct as a single $\gamma$. 

The total detection efficiency of a signature $s$ and a dataset $ds$ is:
\begin{equation}
\label{eq:overall-efficiency}
    \varepsilon_{s, ds} = \varepsilon_{mc} \left( \varepsilon_{cut, ds}\right)^\mathcal{M} \varepsilon_{ac, ds}.
\end{equation}
The analysis cut efficiency is raised to the $\mathcal{M}^{th}$ power because it models channel-related efficiencies and a multiplet is selected only if all of its members pass the selection cuts. 
A summary of the relevant efficiency values for this analysis is reported in Table \ref{tab:scenario-det-efficiency}. 
All the values are weighted by the dataset exposure. 
\begin{table*}[htpb]
\caption{Efficiency terms for all the \nnbEC\ decay signatures. We report exposure weighted averages for both the analysis cut and the anti-coincidence efficiency, that are evaluated on a dataset basis. The analysis cut efficiency is raised to the $\mathcal{M}^{th}$ power. The anti-coincidence efficiency is common to all the signatures. }
\label{tab:scenario-det-efficiency}
  \centering
  \begin{tabular}{ccccc}
   \toprule 
  Signature & Containment & Analysis cut & Anti-coincidence & Total \\\colrule
  (a) & 12.8(5) & 88.7(2) & 99.6(3) & 11.3(5) \\
  (b) & 13.1(5) & 88.7(2) & 99.6(3) & 11.5(5) \\
  (c) & 4.10(20) & 79.0(2) & 99.6(3) & 3.20(13) \\
  (d) & 13.8(6) & 79.0(2) & 99.6(3) & 10.9(4) \\
  (e) & 2.15(9) & 70.5(3) & 99.6(3) & 1.51(6) \\
   \toprule 
  \end{tabular}
\end{table*}
\section{Fit strategy} 
\label{sec:Fit}
We perform a simultaneous Bayesian extended maximum likelihood fit over all the signatures using a multidimensional Probability Density Function (PDF). 
The fit is performed using the BAT software package~\cite{Caldwell:2008fw}, which maps the posterior using a Markov Chain Monte Carlo (MCMC).
A \nnbEC\ candidate event is represented by a set of coincident energy releases $\vec{E} = \left( E_{0}, ... , E_{\mathcal{M}-1}\right)$ 
that match all the requirements of a specific decay signature. 

We define a fit function based on the analysis of the signal signatures from Sec.~\ref{subsec:signatures} and of the background simulations in the selected fit ranges (Tab.~\ref{tab:120TeS_properties}). We generate them from the CUORE Background Model~\cite{Adams:2020bok,Alduino:2017ztm} including the 61 known contaminations of the 
detector setup. The signal is represented by an $\mathcal{M}$-dimensional peak at the energies of Tab.~\ref{tab:120TeS_properties} and background is described as linear in $\mathcal{M}_1$ and uniform for higher multiplicities. 

The fit uses a hybrid of binned fits for $\mathcal{M}_1$ signatures and unbinned fits for $\mathcal{M}_2$ and $\mathcal{M}_3$ signatures. We chose this approach to balance 
the full exploitation of available information with convergence time. In general, unbinned fits are not convenient for highly populated signatures. The likelihood is the product of the five signature likelihoods
\begin{equation}
\mathcal{L}=\prod_{s=(a)}^{(e)}\mathcal{L}_s 
\end{equation}
that we describe in the following.
\subsection{Model for single crystal events}
\label{subsec:M1fit}
For each of the $\mathcal{M}_1$ signatures, and for each dataset, we bin the data into a spectrum of $N_{s,ds}$ bins, with $n_b$ representing the number of events in bin $b$. The likelihood is the product over the datasets $ds$ of Poisson terms:
\begin{equation}
    \mathcal{L}_{s} = \prod^{7}_{ds=1} \ \prod^{N}_{b=1} \ \frac{\mu^{n_b}_b e^{-\mu_b}}{n_b!}
\end{equation}
$\mu_b$ is the expected number of events for bin $b$ -- which is a function of the floating parameters in the fit. 
Since the bin widths $\Delta E_b$ are relatively small, we evaluate $\mu_b$ using point estimates at the bin centers instead of integrals over the bin widths:
\begin{equation}
\label{eq:mu_b}
\mu_b = \mu_{tot} \int^{E^{max}_b}_{E^{min}_b} f(E) \ dE \approx \mu_{tot} \cdot \Delta E_b \cdot f_{s,ds} \left( E_b\right)
\end{equation}
$\mu_{tot}$ is the total expected number of events in the fit range for signature $s$ and dataset $ds$ (Eq.~\ref{eq:mu_tot}), $f_{s,ds} \left( E_b\right)$ is the expected energy distribution of signal and background for the current signature-dataset pair.

We evaluate the detector response function to a monochromatic $\gamma$ peak separately for each crystal in each dataset by fitting the $^{208}$Tl line at 2615 keV in calibration data which is the most prominent peak~\cite{Campani_PhD-thesis:2021,Adams:2020bok}. We account for possible shifts in the reconstructed position of $\gamma$ peaks in the physics spectrum and the energy dependence of the bolometers resolution by including two independent quadratic corrections~\cite{Campani_PhD-thesis:2021,Adams:2020bok}. We define a dataset-dependent rather than channel-dependent correction to model both effects.

We build the binned response function to monochromatic peaks for each dataset by computing the exposure weighted average of all the active channels shapes and evaluating it at the center of each bin in the spectrum:
\begin{equation}
\label{eq:avgBinnedPdf}
 \bar{f}_{ds} \left( E_b \big| \vec{\theta} \right) \propto \sum_{cr} \left(M\Delta T\right)_{cr,ds} \cdot f_{cr,ds}\left(E_b \big| \vec{\theta} \right)
\end{equation}
Here, $(M\Delta T)_{cr,ds}$ denotes the exposure [kg $\cdot$ yr], i.e. the product of the \TeO\ detector mass and measurement live time, for crystal $cr$ in that dataset. $\vec{\theta}$ are the shape parameters to tune the peak position and resolution. 
The constant of proportionality makes $\bar{f}_{ds}\left(E_b \big| \theta \right)$ integrate to 1.

Based on the CUORE background model and \Tecxx\ \nnbEC\ decay simulations, we model the spectrum with a posited signal peak at 1203.8\,keV for signature $(a)$ and 1714.8\,keV for signature $(b)$, a linear background continuum, and additional $\gamma$ peaks that must be included in the fit range.
Thus,
\begin{equation}
\begin{split}
f_{s,ds} \left( E_b\right) &\propto S_{s,ds} \cdot \bar{f}_{ds} \left( E_b \big| \vec{\theta}_{0 \nu} \right) \\
&+ \sum_{i} P_{i,s,ds} \cdot \bar{f}_{ds} \left( E_b \big| \vec{\theta}_{i} \right) \\
&+ B_{s,ds} \cdot \frac{1}{\Delta E_s} \cdot \big[ 1+ m_s (E_b-E^{s}_0)\big]
\end{split}
\end{equation}
where $S_{s,ds}$, $P_{i,s,ds}$ and $B_{s,ds}$ indicate the expected number of \nnbEC\ decays, of events from the i$^{\mathrm{th}}$ residual gamma peak, and of events from a uniform background for the current scenario and dataset, respectively. The constant of proportionality is chosen so that the sum over the bins is unity. $m_s$ describes the slope of the background distribution and $E^s_0$ denotes the center of the ROI for the signature selected. 

The expected number of signal events is 
\begin{equation}
\label{eq:S-norm}
S_{s,ds} = \frac{N_A \eta(^{120}\mathrm{Te})}{m\mathrm{(TeO_2)}} \Gamma_{0 \nu}  \left( M \Delta T \right)_{ds} \varepsilon_{s,ds}
\end{equation}
where $\Gamma_{0 \nu}$ is the signal decay rate, i.e. the parameter of interest, $N_A$ is the Avogadro number, $\eta(^{120}\mathrm{Te})$ is the \Tecxx\ isotopic abundance, $m\mathrm{(TeO_2)}$ is the \TeO\ molecular mass [kg/mol], and $\varepsilon_{s,ds}$ is the total detection efficiency (Eq.~\ref{eq:overall-efficiency} with $\mathcal{M}=1$).

The expected number of events from the i$^{\mathrm{th}}$ residual $\gamma$ peak is
\begin{equation}
    P_{i,s,ds} = P_{i,s} \left(M\Delta T\right)_{ds} \varepsilon_{ac,ds} \varepsilon_{cut,ds}
\end{equation}
where $P_{i,s}$ is the reconstructed amplitude of the gamma line in \cky.

Finally, the expected contribution from continuum background is: 
\begin{equation}
    B_{s,ds} = \mathrm{BI}_s \cdot \Delta E_s \cdot \left( M \Delta T \right)_{ds}
\end{equation} 
where BI$_s$ is the background index for signature $s$ in units of \ckky\ and $\Delta E_s$ indicates the width of the ROI.
Then, the total expected number of events for signature $s$ and dataset $ds$ is
\begin{equation}
\label{eq:mu_tot}
    \mu_{tot} = S_{s,ds} + B_{s,ds} + \sum_{i} P_{i,s,ds} \ .
\end{equation}
The fit parameters are the signal rate $\Gamma_{0\nu}$, the BI$_s$, the $m_s$ and $P_{i,s}$, with the last three included as nuisance and marginalised over.
\subsection{Model for $\mathcal{M}>1$ signatures} 
\label{subsec:M>1fit}
For $\mathcal{M}>1$ signatures, we use an unbinned fit with the following likelihood:
\begin{equation}
    \mathcal{L}_{s} = \prod^{7}_{ds=1} \frac{\left(\lambda_{s,ds}\right)^{n_{s,ds}} e^{-\lambda_{s,ds}}}{n_{s,ds}!} \prod^{n_{s,ds}}_{i=1} f(\vec{E}_{i})
\end{equation}
where $n_{s,ds}$ is the total number of observed events and $\lambda_{s,ds}$ is the expectation value given all possible contributions to the spectrum. $\vec{E}_{i} = (E_0, ... , E_{\mathcal{M}-1})$ is the list of energy depositions for a certain event $i$, and $f(\vec{E}_i)$ is an analythical model for the observed event distribution, which we describe in the following.

\begin{figure*}
    \centering
    \includegraphics[width=0.48\textwidth]{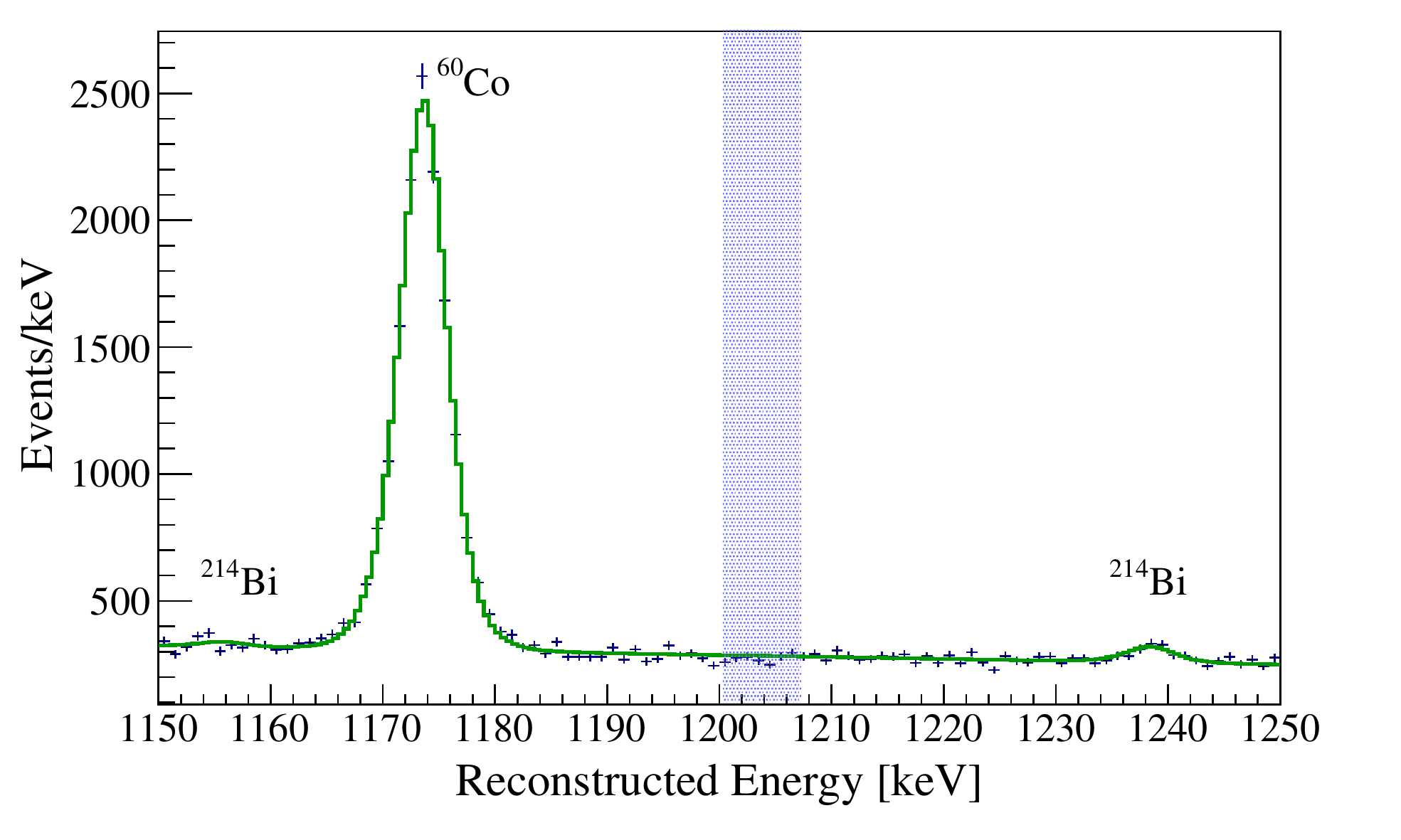} 
    \includegraphics[width=0.48\textwidth]{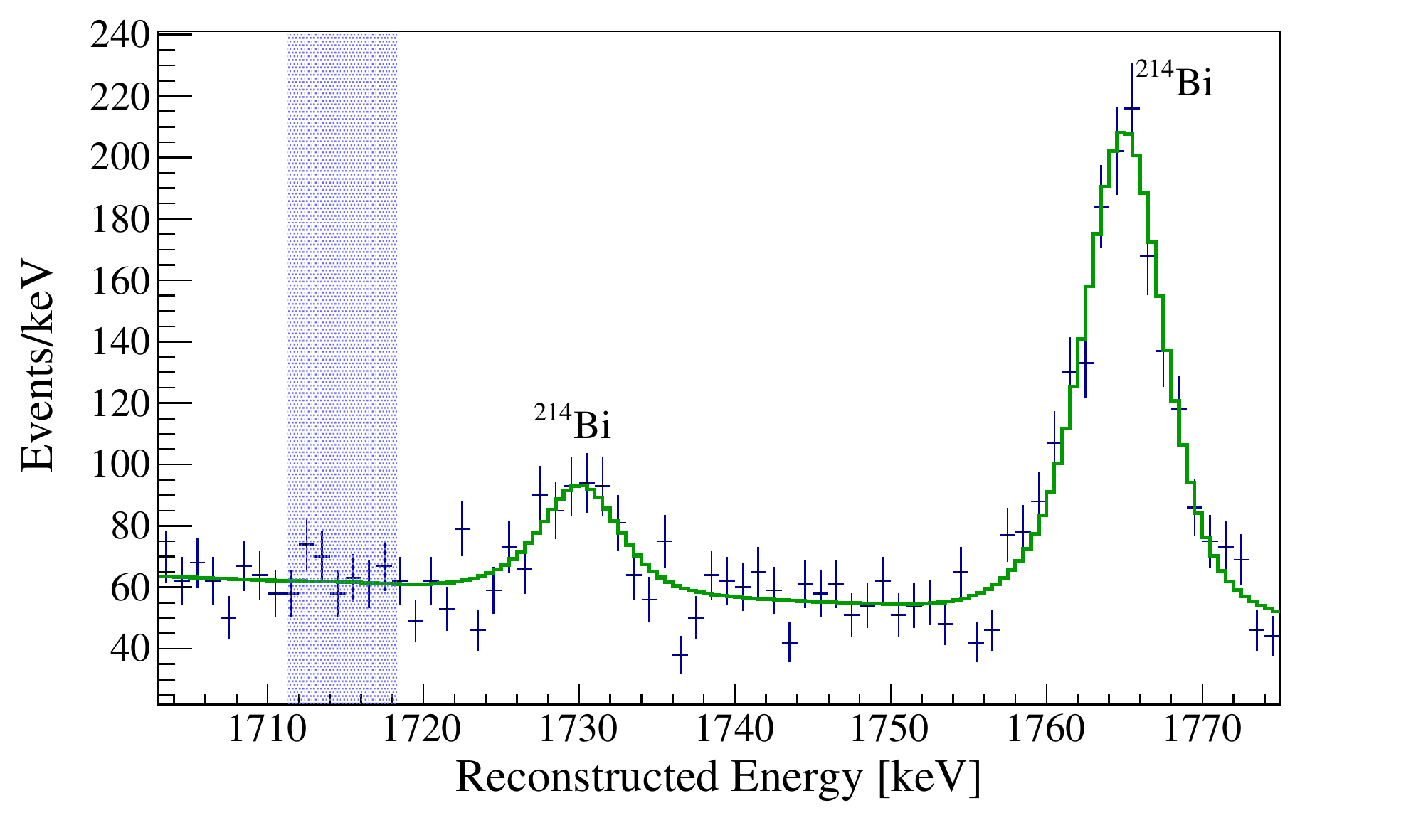}
    \includegraphics[width=0.48\textwidth]{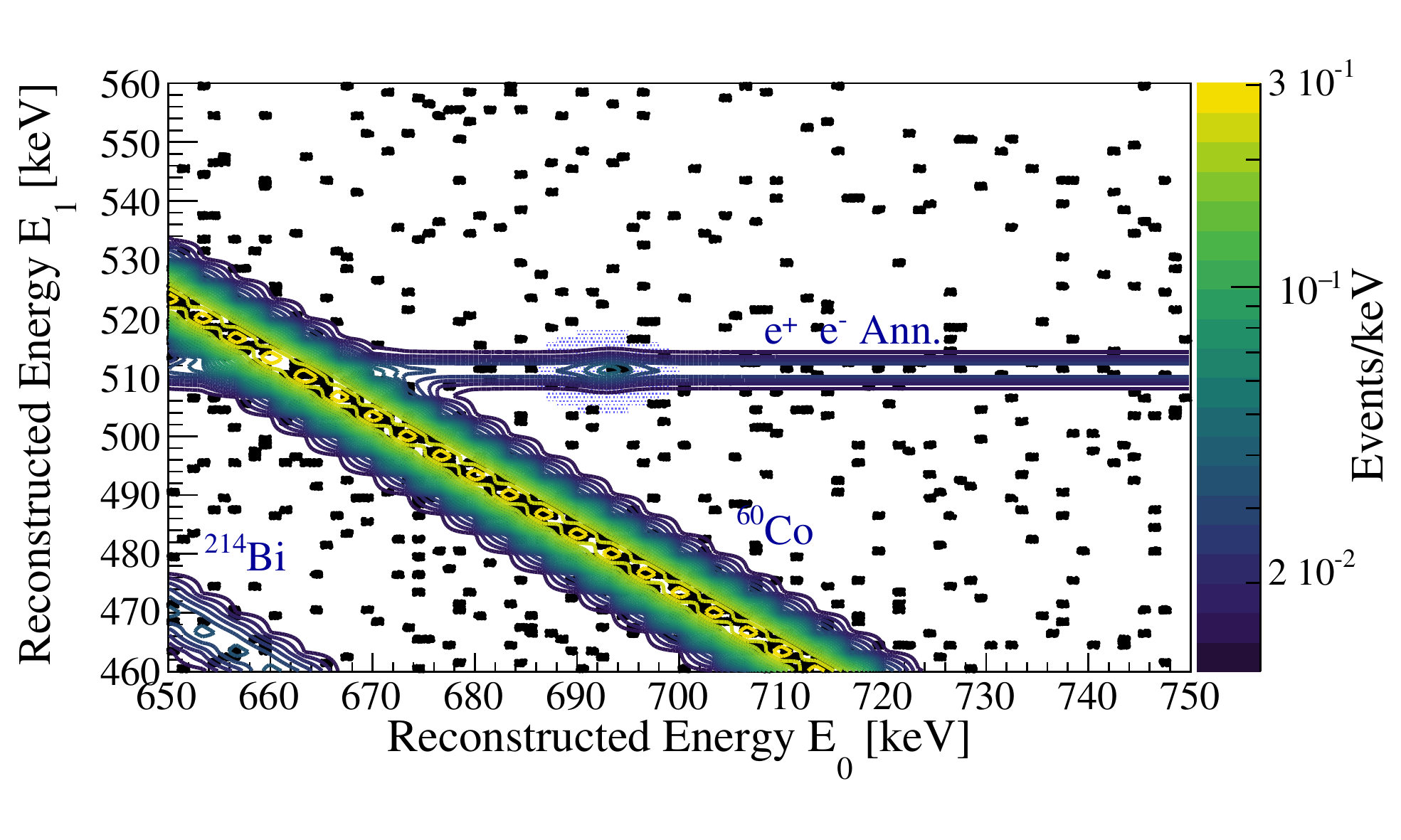}
    \includegraphics[width=0.48\textwidth]{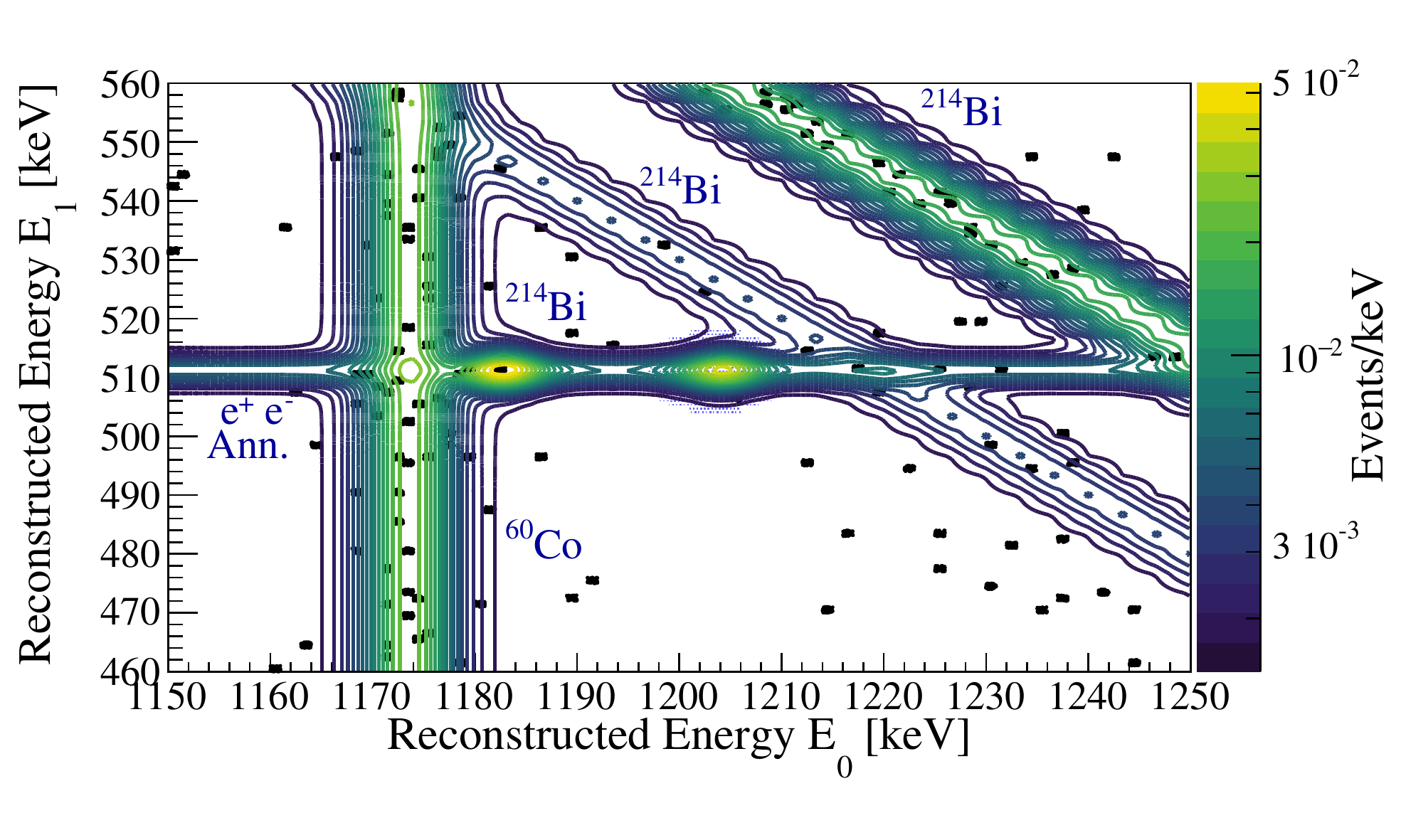}
    \caption{Energy spectrum for signatures (a)-(d). The curves correspond to the best fit minimal model, with the \nnbEC\ decay component normalized to the 90\% CI limit. 
    In the $\mathcal{M}_1$ signatures (top panels) a shaded rectangular area covers a 7 keV region around the expected signal position. In the $\mathcal{M}_2$ signatures (bottom panels), the region whereby the peak is expected is indicated by a shaded circle of radius 7 keV. While the \nnbEC\ decay component is completely hidden by background in signature (a) and (b), the peaks at the 90\% CI limit are clearly visible in signature (c) and especially signature (d). In fact, going from signature (b) to signature (c) the background index decreases by a factor 12 and a factor 100 is gained going from signature (b) to signature (d).
    The labels indicate background peaks and other structures included in the fit model. In the $\mathcal{M}_2$ signatures (bottom panels), the oscillatory pattern of the contour lines is a pure graphical effect. All the plots are done using an exposure weighted average of the crystals response function over each dataset included.} 
    \label{fig:ROI-fits} 
\end{figure*}

Since we are fitting in an $\mathcal{M}$-dimensional space, peaks will appear as $\mathcal{M}$-dimensional energy distributions that we model according to the response function of the set of crystals involved in the event. However, we must also consider Compton scattering for background $\gamma$s, which produce horizontal, vertical and diagonal bands (Fig.~\ref{fig:ROI-fits}). 
From the analysis of CUORE background model we include a peak at (1182,511) keV in scenario (d). This is generated by the 2204.2 keV $\gamma$ from \Bi, that undergoes a pair production followed by an $e^+-e^-$ annihilation. One of the 511 keV $\gamma$s is absorbed in a neighbor crystal while the other escapes undetected. 
Alternatively, if one of the resulting 511\,keV $\gamma$ undergoes Compton scattering within the original crystal and then escapes undetected, while the second 511\,keV $\gamma$ is absorbed in a nearby detector, we measure an $\mathcal{M}_2$ event with 511\,keV in one channel and somewhat less than the single escape peak in the other 
channel. Events of this kind are distributed on horizontal bands (see Fig.~\ref{fig:ROI-fits}). As an example, the 1764.5 keV $\gamma$ emitted by \Bi\ (B.R. $\sim$15\%) can generate events with $E_0\lesssim$1253 keV and $E_1=$ 511 keV that will reconstruct in the spectrum of signature (d). Alternatively, a different background event can consist of two $\gamma$ rays emitted in coincidence.
Signature (d) features a vertical band produced by the full absorption of the 1173\,keV $\gamma$ line from \Co\ on the first channel and a partial energy deposition from the 1332\,keV $\gamma$ on the coincident one (Fig.~\ref{fig:ROI-fits}).
A third possible event configuration is produced by a background $\gamma$ that undergoes Compton scattering in a crystal before being fully absorbed in a neighboring one. The two energy depositions sum to the total energy of the $\gamma$ and the event reconstructs along a diagonal band (Fig.~\ref{fig:ROI-fits}). The distribution of these structures is complicated by the fact that energy depositions on the two crystals are correlated. 
Fortunately, these structures are only relevant for the $\mathcal{M}_2$ signatures.

We define
\begin{equation}
    \lambda_{s,ds} = \lambda_{0 \nu} + \lambda_{bkg} +  \sum_{peak} \lambda_{peak} + \sum_{band} \lambda_{band} + \sum_{diag} \lambda_{diag} 
\end{equation}
where we use the label $band$ and $diag$ to distinguish horizontal/vertical bands and diagonal bands, respectively. For the sake of simplicity, from now on we drop the indexes $s$ and $ds$ if not necessary. We parameterize the expected energy distribution as the sum of five types of distributions:
\begin{equation}
\begin{split}
    f(\vec{E}) &= \frac{\lambda_{S}}{\lambda} f_S(\vec{E}) + \frac{\lambda_{bkg}}{\lambda} f_{bkg}(\vec{E}) + \sum_{peak} \frac{\lambda_{peak}}{\lambda} f_{peak}(\vec{E}) \\
    &+ \sum_{band} \frac{\lambda_{band}}{\lambda} f_{band}(\vec{E}) + \sum_{diag} \frac{\lambda_{diag}}{\lambda} f_{diag}(\vec{E})
\end{split}
\end{equation}
We model the shape of a multi-site signal event as the product of the response functions of the $\mathcal{M}$ detectors involved in the event. 

The distribution of background events depends on the signature. In general, we model it as a linear distribution in $\mathcal{M}$ dimensions:
\begin{equation}
\label{eq:unbinnedBkg}
f_{bkg} \left( \vec{E}_i \right) = \prod_{r = 0}^{M-1} \frac{1}{\Delta E_{r}} \bigg[ 1 + m_{s_r} \big( E_{i_r} - E_{0_r} \big) \bigg]
\end{equation}
where $\Delta E_{r} = E^{max}_r-E^{min}_r$ is the width of the ROI projected along direction $r$ (100 keV for all the signatures) with center $E_{0_r}$. $m_{s_r}$ represents the slope of the background distribution for signature $s$ and direction $r$. The expected number of background events for a given signature-dataset is:
\begin{equation}
    \lambda_{bkg} = \mathrm{BI}_s \left( M \Delta T \right)_{ds} \Delta E_0
\end{equation}
where $\Delta E_{0}$ is chosen as the positron energy range, allowing $\mathrm{BI}_s$ to have units of \ckky.

Background peaks of known $\gamma$ lines are modeled with the same distribution as the signal including the appropriate corrections for energy dependent resolution and reconstruction bias. 
We include a parameter representing the peak intensity $P_{peak}$ whose dimensions are \cky. The total number of expected events for a certain dataset is:
\begin{equation}
    \lambda_{peak} = P_{peak}  (M\Delta T)_{ds} \varepsilon_{ac,ds}  \varepsilon^{\mathcal{M}}_{cut,ds} \ .
\end{equation}
The same holds for horizontal, vertical and diagonal bands, i.e. $\lambda_{band}$ and $\lambda_{diag}$ respectively.

A vertical band consists of a monochromatic peak on the $E_0$ axis and a uniform distribution along the $E_1$ axis. We model the former with the detector response function and the latter with a flat term. The opposite holds for horizontal bands: a uniform energy distribution is included along the $E_0$ axis and a monochromatic peak along the $E_1$ axis.

We describe the shape of diagonal bands in the rotated energy space of $\Sigma=E_0+E_1$ and $\Delta=E_0-E_1$. As with the horizontal and vertical bands, the distribution along the $\Delta$ direction is assumed to be uniform. Along the $\Sigma$ direction, the peak shape is described by the convolution of the detector response functions for the channels involved in the event. 
We account for the energy-dependent resolution functions for each pair of channels, as well as the energy-dependent reconstruction bias. We assume both to be constant across the width of the fit range (100\,keV), which is a good approximation.

All the structures described for $\mathcal{M}_2$ can give rise to additional spectral components in $\mathcal{M}_3$. As an example, \Co\ $\beta$ decay could end up with the 1173-keV $\gamma$ fully absorbed in a crystal (monochromatic peak) and the 1332-keV $\gamma$ making Compton scattering on a nearby bolometer to be finally collected in a third crystal (diagonal band). However, analyzing blinded data we find that none of them produce a significant effect in signature (e). 

We use uniform priors for all the floating statistical parameters, and restrict the range of the amplitude of background components and the \Tecxx\ \nnbEC\ rate to the physical (i.e. non-negative) values.
\subsection{Blinded analysis} 
\label{subsec:blinding}
Before performing the final fit to data, we blind them to validate our fit. We follow an approach similar to~\cite{Adams:2021xns}. 
We inject an unknown, but unrealistically large number of simulated \nnbEC\ decay events into the data. This produces an artificial peak larger than any signal we might expect, that masks the spectral features of the signal region while preserving the background shape and intensity. 
We choose a fake signal rate randomly from the range $[ 6.5 , 30 ]\times10^{-22}\,\mathrm{yr}^{-1}$. This ensures an artificial signal rate larger than the current best 
90\% C.I. upper limit, i.e. $\Gamma_{0\nu} < 2.6 \cdot 10^{-22}$\,yr$^{-1}$~\cite{Alduino:2017dbf}. This choice is justified by the foreseen improvement in terms of sensitivity with CUORE based on the increased exposure, $\sim$10 times that of CUORE-0~\cite{Alduino:2017dbf}, and guarantees that the artificial peak is prominent.

We compute the reconstructed number of counts for each signature and dataset based on their exposures and efficiencies. The actual number of artificial events to be injected is then obtained by Poisson random sampling. Each generated signal event is converted into an appropriate set of $\mathcal{M}$ energy depositions and channels. First, we randomly generate, based on the exposure, the crystal whereby the $\beta^+$ will be absorbed. Then, if present, we select random detectors for the coincident $\gamma$s that satisfy the same radial cut as real data (Sec.~\ref{subsec:CUORE-data}). Finally, we smear the energy deposition of the involved detectors based on their response function centered at the expected peak projection $\mu_r$ (Tab.~\ref{tab:120TeS_properties}) in each direction.

We fit the blinded spectra of each signature including all the possible background structures described in Sec.~\ref{subsec:M1fit} and Sec.~\ref{subsec:M>1fit}. 
Then we remove from the fit model (Tab.~\ref{tab:fitModel120Te}) all the components with negligible significance, i.e. structures for which the lower limit of the 68\% interval around the marginalized mode is zero.  
Finally, as a sanity check, we compare the result of a simultaneous fit on the five signatures with background levels extracted from the fits on single signatures, obtaining compatible values. We employ the intensities extracted from the combined blinded fit (Tab.~\ref{tab:fitModel120Te}) as input for our sensitivity study (Sec.~\ref{subsec:sensitivity}). 

We quantify the fit bias by generating a set of pseudo-experiments with known signal rate, fitting and comparing the resulting rate to the known input rate. Background is generated using the best fit values from the fit on blinded data. We select five evenly spaced values of the \nnbEC\ rate in the range $[ 6.5 , 30 ]\times10^{-22}\,\mathrm{yr}^{-1}$. For each rate, we randomly generate 100 CUORE-like pseudo-experiments. Each of them is made of an ensemble of 7 datasets with the same exposure of the acquired data. We fit each pseudo-experiment and compare the best fit values $\hat{\Gamma}_{0\nu}$ with the injected rates. A linear fit of $\hat{\Gamma}_{0\nu}$ vs $\Gamma^{inj}_{0\nu}$ indicates the absence of a significant bias: 
\begin{equation}
\begin{split}
\hat{\Gamma}_{0\nu} & = p_0 + p_1 \Gamma^{inj}_{0\nu}, \\
p_0 & = (1.1 \pm 0.9) \times 10^{-21} \ \mathrm{yr^{-1}}, \\
p_1 & = (0.992 \pm 0.003). 
\end{split}
\end{equation}
\begin{table*}[htpb]
\centering
\caption{Fit model for \Tecxx\ \nnbEC\ decay analysis and blinded fit result. For each signature, we report the list of all the background structures with the source (isotope) specified. For the horizontal and vertical bands we report the energy corresponding to the full absorption of a background $\gamma$. In the case of diagonal bands we quote the position of the peak in the summed energy, $E_0+E_1$. For each parameter we indicate the best fit value (global posterior mode) extracted from the combined fit on blinded signatures. This was the input of our sensitivity study and fit validation checks. 
}
\label{tab:fitModel120Te}
\begin{tabular}{ccccccccc}
\toprule
& \multicolumn{3}{c}{Energy range [keV]} &  &  & & & \\\cline{2-4}
Signature & $\mathrm{\Delta E_0}$ & $\mathrm{\Delta E_1}$ & $\mathrm{\Delta E_2}$ & Type & Energy (keV) &  Source & Best fit & Units \\\colrule
\multirow{4}{*}{(a)} & \multirow{4}{*}{(1150,1250)} & & & Peak & 1155.2 & \Bi & 0.40(20) & [counts/(kg yr)]\\
& & & & Peak & 1173.2 & \Co & 39.0(4) & [counts/(kg yr)]\\
& & & & Peak & 1238.1 & \Bi & 1.20(20) & [counts/(kg yr)]\\
& & & & $\mathrm{BI_{(a)}}$ & / & / & 0.806(6) & [counts/(keV kg yr)]\\\colrule 
\multirow{3}{*}{(b)} & \multirow{3}{*}{(1703,1775)} & & & Peak & 1729.6 & \Bi & 0.80(10) & [counts/(kg yr)]\\
& & & & Peak & 1764.5 & \Bi & 3.50(16) & [counts/(kg yr)]\\
& & & & $\mathrm{BI_{(b)}}$ & / & / & 0.160(3) & [counts/(keV kg yr)]\\\colrule 
\multirow{4}{*}{(c)} & \multirow{4}{*}{(650,750)} & \multirow{4}{*}{(460,560)}& & Horizontal band & 511 &  / & 0.09(3) & [counts/(kg yr)]\\
& & & & Diagonal band & 1173.2 & \Co & 1.77(9) & [counts/(kg yr)]\\
& & & & Diagonal band & 1120.3 & \Bi & 0.030(13) & [counts/(kg yr)]\\
& & & & $\mathrm{BI_{(c)}}$ & / & / & 0.0139(7)  & [counts/(keV kg yr)]\\\colrule
\multirow{6}{*}{(d)} & \multirow{6}{*}{(1150,1250)} & \multirow{6}{*}{(460,560)} & & Peak & (1182.2,511) & \Bi & 0.015(8) & [counts/(kg yr)]\\
& & & & Horizontal band & 511 &  / & 0.030(20) & [counts/(kg yr)]\\
& & & & Vertical band & 1173.2 &  \Co & 0.19(3) & [counts/(kg yr)]\\
& & & & Diagonal band & 1729.6 & \Bi & 0.020(13) & [counts/(kg yr)]\\
& & & & Diagonal band & 1764.5 & \Bi & 0.090(20) & [counts/(kg yr)]\\
& & & & $\mathrm{BI_{(d)}}$ & / & / & 0.00160(24) & [counts/(keV kg yr)]\\\colrule
(e) & (650,750) & (460,560) & (460,560) & $\mathrm{BI_{(e)}}$  & / & / & 0.00011(5) & [counts/(keV kg yr)] \\
\toprule
\end{tabular}
\end{table*}

\subsection{Exclusion sensitivity} 
\label{subsec:sensitivity}
\begin{figure}[htpb]
    \centering
    \includegraphics[width=0.5\textwidth]{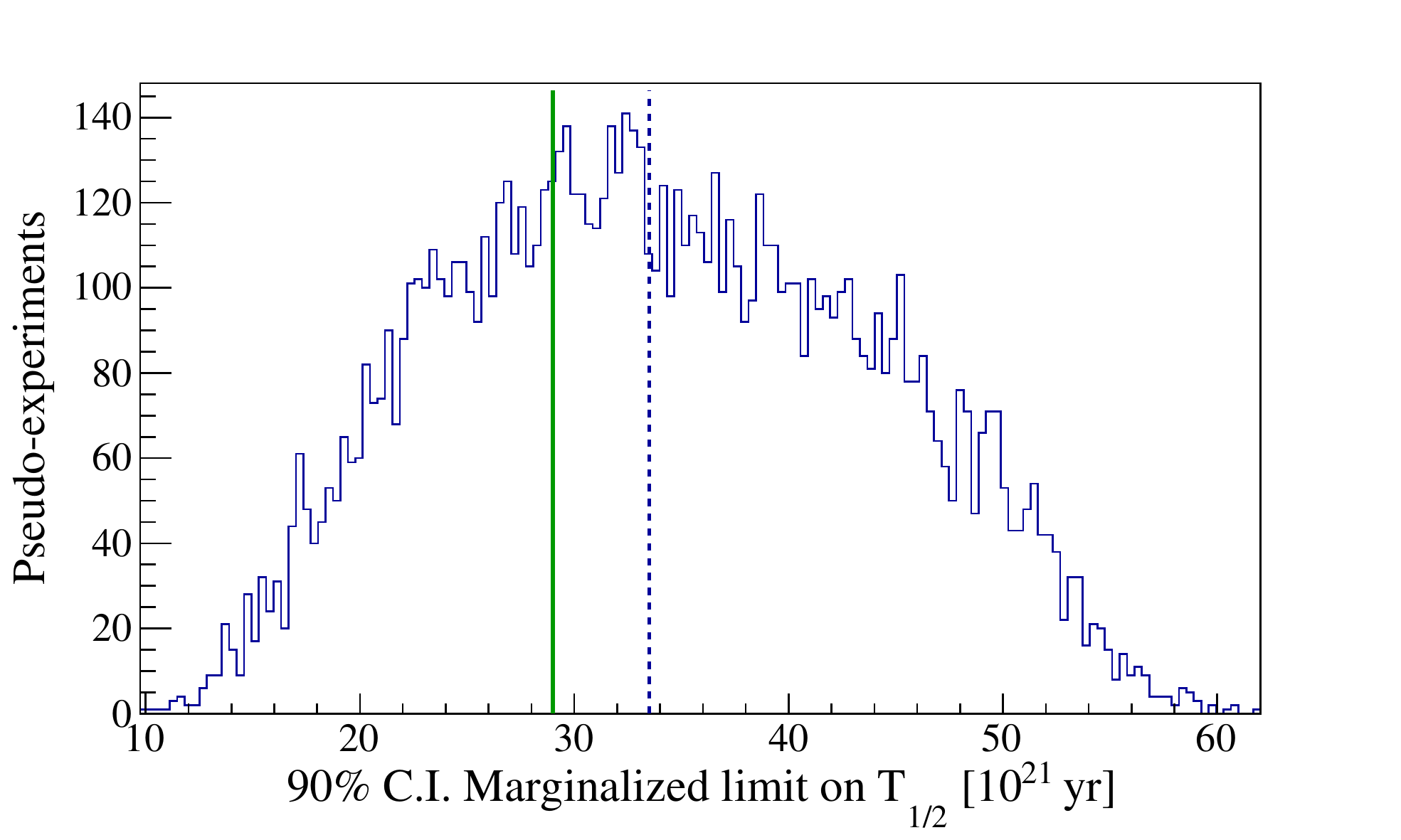}
    \caption{Distribution of the 90\% C.I. marginalized limits on the half-life of \nnbEC\ decay as obtained from pseudo-experiments (Sec.~\ref{subsec:sensitivity}). The median limit setting sensitivity is $S^{0\nu}_{1/2} = 3.4 \cdot 10^{22}$ yr (dashed line). The 90\% C.I. limit from this analysis is shown for comparison (solid line).}
    \label{fig:sensitivity}
\end{figure}

To compute the exclusion sensitivity, we generate 10$^4$ pseudo-experiments populated with only background components, using the intensities reported in Tab~\ref{tab:fitModel120Te}. 
We then fit each pseudo-experiment with the signal plus background model. Finally, we compute the lower limit for \Tecxx\ \nnbEC\ decay half-life from the 90\% quantile of the marginalized posterior distribution for the decay rate. The distribution of such limits is shown in Fig.~\ref{fig:sensitivity}. We obtain a median 90\% C.I. limit setting sensitivity on the half-life of $3.4 \cdot 10^{22}$ yr.

\section{Results}
\label{sec:Results}
\begin{figure}[htpb]
    \centering
    \includegraphics[width=0.5\textwidth]{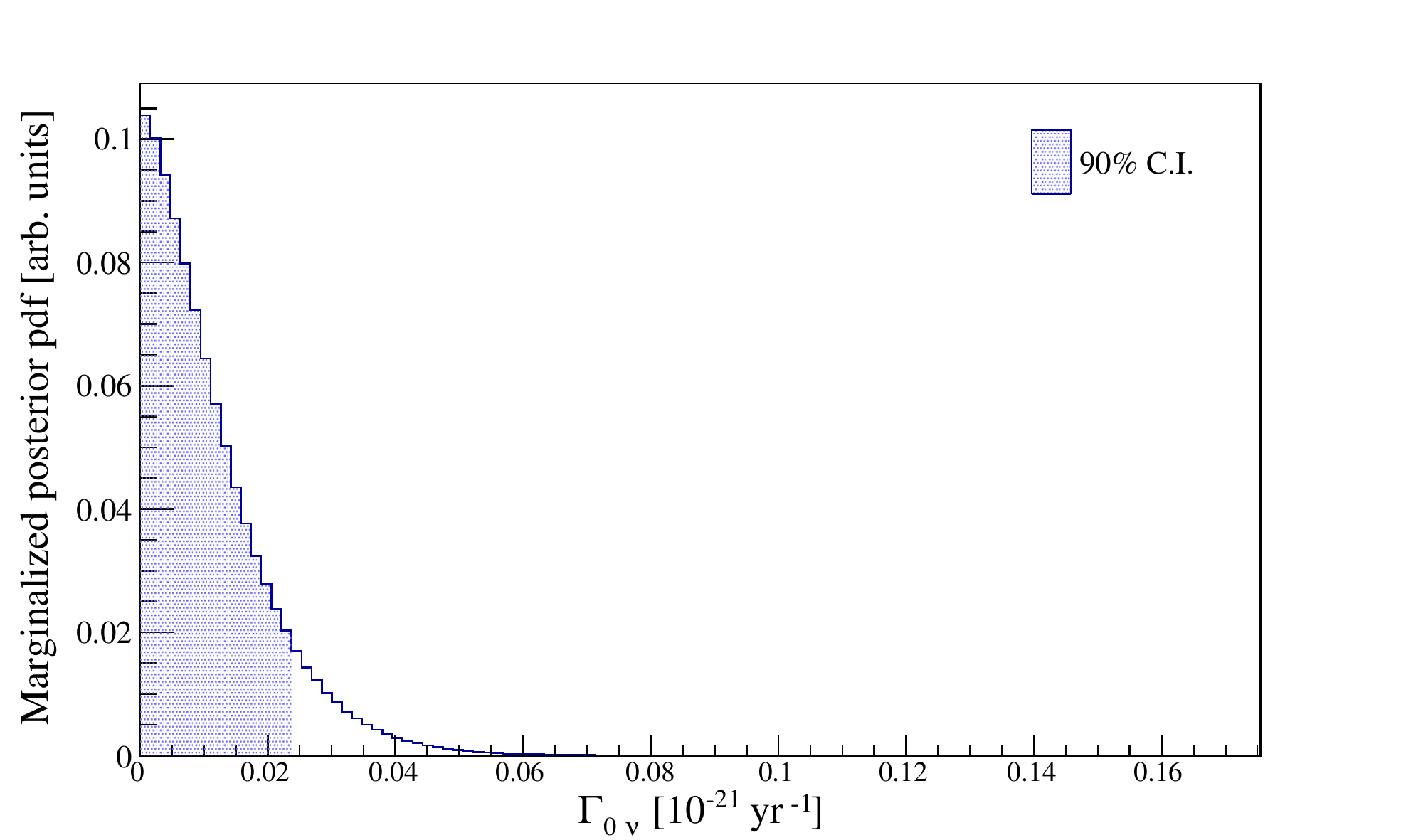}
    \caption{Marginalized \nnbEC\ decay rate posterior PDF from the combined fit including the dominant sources of systematics, i.e. the efficiency terms, \Tecxx\ isotopic composition and the uncertainty on the decay Q-value. The shaded area indicates the 90\% C.I. limit.}
    \label{fig:rate-posterior}
\end{figure}

The result of the combined fit to the unblinded data is shown in Fig.~\ref{fig:ROI-fits}. We find only 4 events matching the requirements of signature (e), not shown in Fig.~\ref{fig:ROI-fits}. 
We find no evidence for neutrinoless $\beta^+EC$ decay in \Tecxx. 
Including contributions from the dominant sources of systematic uncertainty, i.e. \textit{Nuisance Parameters} from  Tab.~\ref{tab:systematics} and uncertainty on the decay Q-value, the global mode (best fit) of the joint posterior distribution for the signal rate is 
\begin{equation} 
    \hat{\Gamma}_{0\nu} = 0.1^{+1.4}_{-0.1} \cdot 10^{-23} \ \mathrm{yr}^{-1}
\end{equation}
where we quote the uncertainty extracted from the smallest 68\% interval around the mode of the $\Gamma_{0\nu}$ marginalized distribution.
We obtain the following lower bound on the \Tecxx\ half-life for \nnbEC\ decay:
\begin{equation}
T^{0\nu}_{1/2} > 2.9 \cdot 10^{22} \ \mathrm{yr} \ (90\% \ \mathrm{C.I.})
\end{equation}
The marginalized posterior PDF is shown in Fig.~\ref{fig:rate-posterior}.

Repeating the fit with non physical values of $\Gamma_{0\nu}$ allowed, we find background parameters consistent with those from the fit with the $\Gamma_{0\nu} \geq 0$ constraint and no under-fluctuation of the signal rate: in this case the best fit signal rate is $\hat{\Gamma}_{0\nu} = 0.2^{+1.0}_{-1.0} \cdot 10^{-23}$ yr$^{-1}$. 
This result is compatible with our exclusion sensitivity, in fact, the limit is looser than expected from the median half-life $S^{0\nu}_{1/2}$, and the probability to obtain a stronger limit is 67.5\%. 

\begin{table*}[htpb]
    \centering
    \caption{Systematic uncertainties affecting the \nnbEC\ decay analysis. 
    Analysis Cut Efficiency I corresponds to base cut efficiency and Analysis Efficiency II indicates the additional effect related to the pulse shape. 
    }
    \begin{tabular}{lcc}
    \toprule
    \multicolumn{3}{c}{Fit parameter systematics} \\\colrule
    Systematic & Prior & Effect on $\Gamma_{0 \nu}$ \\\colrule
    Intrinsic BAT Uncertainty & & 0.4\% \\\colrule 
    \multicolumn{3}{c}{Nuisance Parameters} \\\colrule 
    Analysis Cut Efficiency I & Gaussian & 0.4\% \\
    Analysis Cut Efficiency II & Uniform & 0.2\% \\ 
    Anti-coincidence Veto Efficiency & Gaussian & 0.4\% \\ 
    \nnbEC\ Containment Efficiency & Gaussian & 0.3\%\\
    \Tecxx\ Isotopic Abundance & Gaussian & 2.9\%\\ 
    All Nuisance Parameters Combined & Multivariate & 4.4\% \\\colrule
    \multicolumn{3}{c}{Additional Parameters} \\\colrule
    Energy Scale Bias & Multivariate & 0.2\% \\
    Energy Scale Resolution & Multivariate & 0.1\% \\ 
    \Tecxx\ Q-value & Gaussian & 4.4\% \\ 
    \toprule
    \end{tabular}
        \label{tab:systematics}
\end{table*}

\subsection{Systematic effects}
\label{subsec:syst-fit}
The sources of systematic uncertainty included in this analysis are given in Table~\ref{tab:systematics}. We include contributions related to the efficiencies, either data-driven or a result of Monte Carlo simulations, the uncertainty on the \Tecxx\ isotopic composition $\eta(^{120}\mathrm{Te})$~\cite{Meija:2016IUPAC} and on the decay Q-value~\cite{Scielzo:2009nh}, and the energy dependence of the detector response function. Finally, there is the intrinsic uncertainty induced by the Bayesian (MCMC) fitting procedure. We evaluate the latter by repeating the minimal model fit, i.e. no systematics, under the same conditions 10$^3$ times and fitting the resulting distribution of the 90\% C.I. limits on the signal rate with a Gaussian. We obtain a value of $\Gamma^{0\nu}_{90} = (2.430 \pm 0.010) \cdot 10^{-23}$ yr$^{-1}$, yielding a 0.4\% relative effect on our final result. 

We split the remaining systematics into two categories, which we estimate with two different approaches. We refer to the first set as \emph{Nuisance Parameters} and treat them as additional parameters in the fit with associated priors, which are then marginalized over. These systematics include the uncertainty on the isotopic abundance $\eta(^{120}\mathrm{Te})$, which we treat as a single global parameter with a Gaussian prior, and the uncertainty on the PSA cut efficiency, which is treated as a single global parameter with a uniform prior. We also include uncertainties on the base cut and anti-coincidence cut efficiencies, which are treated at the per-dataset level with Gaussian priors. Since the containment efficiency is set at the signature level, this contribution accounts for 5 additional parameters with Gaussian priors. Table~\ref{tab:systematics} lists the effect of each of these sources of uncertainty when treated independently, as well as their combined effect. The majority of them have minimal effect on the final result, consistent with the intrinsic uncertainty induced by the BAT fit. The leading source of uncertainty is the 11\% uncertainty on the isotopic abundance, which translates directly to a 2.9\% variation of the half-life limit. The combined effect from these Nuisance Parameters is 4.4\%.

For the second set of systematics, we were not able to employ a fully Bayesian approach due to computational limitations. We refer to these parameters as the \emph{Additional Parameters}. They include the uncertainty on the \nnbEC\ decay Q-value, and the uncertainties associated with the detector response function. 
We manually marginalize over these parameters at a discrete set of points. For the uncertainty on the \nnbEC\ decay Q-value, we sample 11 points spanning the $\pm3\sigma$ range about the central value. Similarly, for both the uncertainty on the detector energy resolution and on the reconstruction bias as a function of energy we sample 1331 points -- 11 points along each of the 3 dimensions that define the energy dependence. As shown in Table~\ref{tab:systematics}, the corrections on the peak position and the energy resolution induce effects $\lesssim0.2$\%, while the uncertainty on the \nnbEC\ decay Q-value affects the limit by 4.4\%. 
In this case we obtain a smaller value of the limit on the signal rate. This result is driven by an under-fluctuation of the continuum background below 1203.8~keV in signature (d), that is the most sensitive one based on containment efficiency (Tab.~\ref{tab:120TeS_properties}) and background index (Tab.~\ref{tab:fitModel120Te}). 

In conclusion, the dominant systematics are the combined effect of the efficiencies and \Tecxx\ isotopic composition and the uncertainty on the decay Q-value. We combine them together to extract the final limit on \nnbEC\ decay of \Tecxx\ by simultaneously sampling 11 points in the space of the Q-value and letting the \textit{Nuisance Parameters} from Tab.~\ref{tab:systematics} float in each fit. We obtain a lower bound of $T^{0\nu}_{1/2} > 2.9 \cdot 10^{22}$ yr (90\% C.I.) on \Tecxx\ half-life for such decay.

\section{Conclusions}
\label{sec:finalconclusions}
We have presented the latest search for neutrinoless $\beta^+EC$ decay of \Tecxx\ with CUORE based on seven datasets
corresponding to a 355.7 kg$\cdot$yr \TeO\ exposure and a 0.2405 kg$\cdot$yr \Tecxx\ exposure. We found no evidence for such transition and placed a Bayesian lower limit on the decay half-life of $T^{0\nu}_{1/2} > 2.9 \cdot 10^{22}$ yr at 90\% C.I., including the dominant systematic uncertainties. 
Considering our median exclusion sensitivity $S^{0\nu}_{1/2} = 3.4 \cdot 10^{22}$ yr, the probability to obtain a stronger limit is 67.5\%.

This result represents the most stringent limit to date on \nnbEC\ decay of \Tecxx, yielding a factor 10 improvement with respect to the combined analysis of CUORE-0 and Cuoricino~\cite{Alduino:2017dbf}. This is a consequence of the increased exposure and larger containment efficiency, mostly due to the higher capability of detecting the 511\,keV gammas in active parts of the detector. Furthermore, it proves the effectiveness of the coincidence analysis based on the CUORE detector modularity for rare decay searches. 

Further improvements in the sensitivity to \nnbEC\ decay could result from the inclusion of not-fully contained events, that would increase our detection efficiency, and from a study of the geometrical distribution of candidate signal events as opposed to background. 

 \begin{acknowledgments}
 The CUORE Collaboration thanks the directors and staff of the Laboratori Nazionali del Gran Sasso and the technical staff of our laboratories. This work was supported by the Istituto Nazionale di Fisica Nucleare (INFN); the National Science Foundation under Grant Nos. NSF-PHY-0605119, NSF-PHY-0500337, NSF-PHY-0855314, NSF-PHY-0902171, NSF-PHY-0969852, NSF-PHY-1307204, NSF-PHY-1314881, NSF-PHY-1401832, and NSF-PHY-1913374; and Yale University. This material is also based upon work supported by the US Department of Energy (DOE) Office of Science under Contract Nos. DE-AC02-05CH11231 and DE-AC52-07NA27344; by the DOE Office of Science, Office of Nuclear Physics under Contract Nos. DE-FG02-08ER41551, DE-FG03-00ER41138, DE-SC0012654, DE-SC0020423, DE-SC0019316; and by the EU Horizon2020 research and innovation program under the Marie Sklodowska-Curie Grant Agreement No. 754496. This research used resources of the National Energy Research Scientific Computing Center (NERSC). This work makes use of both the DIANA data analysis and APOLLO data acquisition software packages, which were developed by the Cuoricino, CUORE, LUCIFER and CUPID-0 Collaborations.

 \end{acknowledgments}
 \bibliography{bibliography}

\end{document}